\documentclass[showpacs,superscriptaddress,floatfix,reprint,prb]{revtex4-1}
\usepackage{graphicx}
\usepackage{amsmath}
\usepackage{amssymb}
\usepackage{braket}

\begin{document}

\title{Fermion-parity switches of the ground state of Majorana billiards}
\author{Bar{\i}\c{s} Pekerten}
\author{A. Mert Bozkurt}
\author{\.{I}nan\c{c} Adagideli}
%\email{adagideli@sabanciuniv.edu}
\affiliation{Faculty of Engineering and Natural Sciences, Sabanci
University, Orhanl{\i}-Tuzla, 34956 \.{I}stanbul, Turkey}

\date{\today}

\begin{abstract}
Majorana billiards are finitely sized, arbitrarily shaped
superconducting islands that host Majorana bound states. We study the
fermion-parity switches of the ground state of Majorana billiards. In
particular, we study the density and statistics of these fermion-parity
switches as a function of applied magnetic field and chemical
potential. We derive formulae that specify how the average density of
fermion-parity switches depends on the geometrical shape the billiard.
Moreover, we show how oscillations around this average value is
determined by the classical periodic orbits of the billiard. Finally,
we find that the statistics of the spacings of these fermion-parity
switches are universal and are described by a random matrix ensemble,
the choice of which depends on the antiunitary symmetries of the system
in its normal state. We thus demonstrate that ``one can hear
(information about) the shape of a Majorana billiard'' by investigating
its ``fermion-parity switch spectrum''.
\end{abstract}


% 73.22.-f	Electronic structure of nanoscale materials and related systems
% 74.78.Na 	Mesoscopic and nanoscale systems
% 74.20.Mn 	Nonconventional mechanisms
% 71.23.-k 	Electronic structure of disordered solids

\maketitle

%%%%%%%%%%%%%%%%%%%%%%%%%%%%%%%%%%%%%
%									%
%	Introduction					%
%									%
%%%%%%%%%%%%%%%%%%%%%%%%%%%%%%%%%%%%%

\section{Introduction}\label{SECT:Introduction}

Eigenvalue spectra of finite quantum systems are related to their shape
in the short wavelength limit~\cite{REF:Book:Weyl68,
REF:Book:Baltes76}. The celebrated Weyl expansion relates the smooth
part of the density of states to the volume, boundary area, curvature
as well as the Euler characteristics of the shape of the
system~\cite{REF:Balian70, REF:Book:Baltes76, REF:Book:Brack03}. The
remaining part, namely the density of states fluctuations, sensitively
depends on the set of periodic orbits of the corresponding classical
dynamics as well as the type of scattering featured in the
system~\cite{REF:Balian72, REF:Berry85, REF:Book:Gutzwiller90,
REF:Book:Mehta04, REF:Wurm09}. Moreover, if all unitary symmetries are
completely broken, the level-spacing distribution becomes universal and
reflects the presence (or absence) of antiunitary
symmetries~\cite{REF:Wigner55, REF:Dyson62, REF:Dyson62b, REF:Dyson63,
REF:Bohigas84, REF:Book:Mehta04}.

The ground state of conventional superconductors have even number of
fermions, reflecting their completely paired nature
(\textit{even fermion-parity}). However, under
certain conditions, {the energy level of} a state with an
odd number of fermions (\textit{odd
fermion-parity}) can cross the the energy level
of the state with even fermion-parity to become the new
ground state. This crossing, dubbed fermion-parity crossing (FPX), is
protected since perturbations that mix different fermion-parity states
are prohibited. While well known within the context of impurity states
in superconductors~\cite{REF:Balatsky06, REF:Mi14}, these crossings can
be viewed as topological phase transitions~\cite{REF:Fu09, REF:Ryu10,
REF:Stanescu11, REF:Lee12, REF:Beenakker13b, REF:Rieder13, REF:Chang13,
REF:Sau13b, REF:Badiane13, REF:Chevallier2013, REF:Lee14, REF:Hedge15}.
The modes that form at the
degeneracy point are the well known Majorana
zero modes featuring non-Abelian statistics~\cite{REF:Kitaev01,
REF:Hasan10, REF:Qi11, REF:Alicea12, REF:Book:Bernevig13,
REF:Elliott15}, which have attracted recent attention as the
candidate system for realization of topological quantum computers.

Currently there are experimental signatures of zero-bias conductance
peaks, suggestive of edge-bound zero-bias states~\cite{REF:Mourik12,
REF:Nadj-Perge14, REF:Fornieri19, REF:Vaitiekenas18}. However,
conclusive experimental demonstration of the Majorana bound states has
been elusive so far as these observed peaks could have non-topological
origins such as Andreev bound states~\cite{REF:Lee14, REF:Hsieh2012,
REF:Prada12, REF:Chevallier2013, REF:Silva2016, REF:Liu2017,
REF:Nichele2017, REF:Zuo2017, REF:Tang2018, REF:Moore2018a,
REF:Hell2018, REF:Liu18, REF:Vuik2018, REF:Moore2018b, REF:Reeg2018,
REF:Kayyalha2019, REF:Chen2019, REF:Woods2019, REF:Cao2019}, Kondo
effect, weak antilocalization, and disorder~\cite{REF:Motrunich2001,
REF:Brouwer2011a, REF:Brouwer2011b, REF:Pikulin12, REF:Popinciuc2012,
REF:Bagrets12, REF:Liu12, REF:Neven2013, REF:Churchill13, REF:Sau2013a,
REF:Pan2019}. Hence new methods of distinguishing Majorana zero modes
from other sources as well as new ways of understanding these nanowires
has become desirable. The presence of FPX sequences has been regarded
as the smoking gun signature of Majorana states in ballistic 1D
wires~\cite{REF:DasSarma12, REF:Rodriguez-Mota19}. The universal
statistics of these FPXs were first studied by Beenakker \textit{et
al}~\cite{REF:Beenakker13b}. Recent measurements on proximity coupled
nanowires, expected to feature topological superconductivity, found
sequences of FPXs as a function of magnetic field as well as gate
voltage~\cite{REF:Chen2019, REF:Woods2019}.

In this work, we study the FPXs in finite sized topological
superconducting systems through the lens of (i)~spectral geometry,
(ii)~semiclassical physics and (iii)~random matrix theory. We call
these finite superconducting systems that feature FPXs \textit{Majorana
billiards} (MBs)~\cite{REF:Footnote:MBRealizations}. These FPXs in MBs
occur as an external parameter of the system, such as the chemical
potential $\mu$ or the Zeeman energy $B$, is varied. We call the set of
values at which FPXs occur \textit{(FPX) spectrum}, and the elements of
this set \textit{FPX points}. We first extract geometrical information
from the FPX spectrum. In particular, we investigate the relation
between the average density of FPXs and the geometry of the system. In
other words, we ask and answer the question whether one can ``hear''
the shape of a Majorana billiard from its FPX spectrum, alluding to
Kac's famous question (as phrased by L.~Bers), ``Can one hear the shape
of a drum?''~\cite{REF:Kac66, REF:Footnote:IsospectralDomains}. In the
same spirit, we next explore the connection between the dynamics of MBs
and the oscillations around the average density of FPXs. These
oscillations are analogous to supershell effects in nuclei, atomic
clusters or nanoparticles~\cite{REF:Book:Brack03}. To the best of our
knowledge, there has been no theoretical investigation of these
supershell effects in MBs so far. We stress that as the FPX spectrum is
experimentally accessible~\cite{REF:Shen2018, REF:Chen2019}, it would
be possible to analyze available experimental data on FPXs and observe
the shell and supershell effects predicted in this manuscript. Finally,
we show that the FPX spectrum of MBs exhibits universal statistics that
depends on whether the underlying normal system is regular, diffusive,
chaotic or localized.

Our manuscript is organized as follows: In 
Section~\ref{SECT:SystemDescription}, we describe the physical systems 
that we focus on in this work. In Section~\ref{SECT:PXingForMB}, we 
focus on the average density of FPXs of a MB and study the relation 
between this density and the geometry of a MB billiard. In addition, we 
derive a scaling property of FPX points for a spinful Majorana 
billiard. We also show how non-zero density of FPX points in disordered 
systems are induced below the clean-system topological phase 
transition, analogous to Lifshitz tails in disordered systems. In 
Section~\ref{SECT:DOSOscillations}, we discuss the oscillatory part of 
the density of FPXs due to supershell effects and how it relates to 
classical periodic orbits of the billiard. In 
Section~\ref{SECT:UniversalStats}, we focus on the universality of the 
statistics of FPXs in integrable and chaotic MBs and explore the 
universality crossover as the system goes from diffusive to localized.

%%%%%%%%%%%%%%%%%%%%%%%%%%%%%%%%%%%%%
%									%
%	System description				%
%									%
%%%%%%%%%%%%%%%%%%%%%%%%%%%%%%%%%%%%%

\section{Description of the system}\label{SECT:SystemDescription}

\subsection{Majorana Billiards from \textit{s}- and \textit{p}-wave topological superconductors}\label{SUBSECT:SPWaveMB}

%%%%%%%%%%%%%%%%%%%%%%%%%%%%%%%%%%%%%
%									%
%	Hamiltonians					%
%									%
%%%%%%%%%%%%%%%%%%%%%%%%%%%%%%%%%%%%%

\begin{figure}[tb]
\includegraphics[width=\columnwidth]{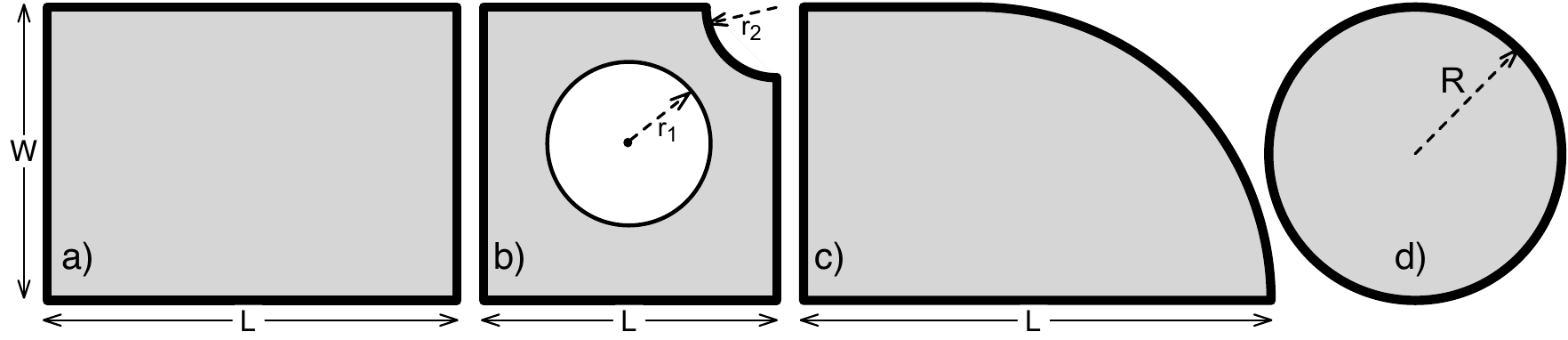}
\caption{The 2D geometries used in the tight-binding numerical
simulations: a) Rectangle, b) Lorentz gas cavity, c) Quarter-stadium
cavity, d) Disk.}\label{FIG:2DSystemShape}
\end{figure}

We study finite 2D Majorana billiard systems whose dynamics are
described by the Bogoliubov--de Gennes
Hamiltonian~\cite{REF:deGennes99}
\begin{align}\label{EQN:Hamiltonian_SWave}
H_s &= h(\mathbf{p},\mathbf{r})\, \tau_z+\alpha (p_x \sigma_y-p_y \sigma_x)  \tau_z + B \sigma_x + \Delta \tau_x,
\end{align}
where $\sigma_i$ [$\tau_i$] are the Pauli matrices in spin
[particle-hole] space ($i=x,y,z$), $h(\mathbf{p},\mathbf{r})=p^2/2m +
V(\mathbf{r})-\mu$ is the spinless part of the single-particle
Hamiltonian with $\mu$ being the chemical potential, $\alpha$ is the
Rashba spin-orbit coupling strength, $B$ is the Zeeman energy and
$\Delta$ is the \textit{s}-wave pair potential and $V({\bf r})$ is the
single-particle potential which consists of disorder and confinement
potentials. The systems can be clean or disordered, and their dynamics
can therefore be ballistic chaotic/integrable or diffusive in the
classical limit. Hence our numerical tight-binding simulations focus on
these cases as shown in  Fig.~\ref{FIG:2DSystemShape}.

For a one dimensional system, if the Zeeman
energy is large enough to deplete one of the spin-polarized bands of
the Hamiltonian in Eq.~(\ref{EQN:Hamiltonian_SWave}), the system is
described by a spinless Bogoliubov--de Gennes Hamiltonian with an
effective p-wave pair potential~\cite{REF:Lutchyn10,REF:Oreg10}.
In this work, we consider this system as well as its 2D
generalization, whose Hamiltonian is given by
\begin{equation}\label{EQN:Hamiltonian_PWave}
H_p = h(\mathbf{p},\mathbf{r})\,\tau_z + \Delta' \boldsymbol{\tau}\cdot\mathbf{p},
\end{equation}
where $\Delta' = \alpha\, \Delta / \epsilon$ is the (p-wave) pair
potential strength, with $\epsilon = \sqrt{B^2-\Delta^2}$ for
$B>\Delta$. Throughout this manuscript, we call systems featuring the
Hamiltonian $H_s$ [$H_p$] ``\textit{s}-wave'' [``\textit{p-wave}''].

%%%%%%%%%%%%%%%%%%%%%%%%%%%%%%%%%%%%%
%									%
%	Figure 1: Shapes				%
%									%
%%%%%%%%%%%%%%%%%%%%%%%%%%%%%%%%%%%%%

%%%%%%%%%%%%%%%%%%%%%%%%%%%%%%%%%%%%%
%									%
%	Problem description				%
%									%
%%%%%%%%%%%%%%%%%%%%%%%%%%%%%%%%%%%%%

\subsection{Density of fermion-parity crossings}\label{SUBSECT:ProblemDescription}

We now define the density of fermion-parity crossings. We envision
finding the zero energy solutions of $H_s$ and $H_p$ in
Eqs.~(\ref{EQN:Hamiltonian_SWave}) and
(\ref{EQN:Hamiltonian_PWave}) as an external parameter is varied. This
parameter for $H_p$ is the chemical potential $\mu$. For $H_s$, the
external parameter could either be the chemical potential $\mu$ or the
Zeeman energy $B$. We then record the values of these parameters at which
$H_s$ or $H_p$ have zero energy solutions as the FPX points. (We show
below in Section~\ref{SUBSECT:SWaveUniversal} that the FPX points of a
given \textit{s}-wave MB with respect to $\mu$ and with respect to $B$
are related.)
Finally we define the density of FPX points of a MB with
respect to the dimensionless parameter $\beta$ ($\beta = \mu/t$ or
$\beta = B/t$) as
\begin{align}\label{EQN:DoPXings}
\rho(\beta) &\equiv \sum_i \delta(\beta-\beta_i),
\end{align}
where $\beta_i = \mu_i/t$ or $\beta_i = B_i/t$, $\mu_i$ and $B_i$ are
the FPX points and $t$ determines the the bandwidth of the system in
that in in $d$ dimensions the bandwidth is $2dt$. (In tight-binding
simulations, $t=\hbar^2/2ma^2$ is the hopping term and $a$ is the
lattice parameter.) We also define the integrated density
$\mathcal{N}(\beta)$ of FPX points, given by
\begin{align}\label{EQN:IntegratedDoPXings}
\mathcal{N}(\beta) &= \int_{-\infty}^\beta \rho(\beta') \, d\beta'.
\end{align}

We separate the density $\rho(\beta)$ into its average value
$\bar{\rho}(\beta)$ and the oscillations around this average $\rho_{\rm
osc}(\beta)$ as is customary in the semiclassical study of the DOS of a
billiard~\cite{REF:Balian70, REF:Book:Brack03, REF:Balian72,
REF:Berry85,REF:Book:Gutzwiller90} and write
\begin{align}\label{EQN:DOSMeanPlusOsc}
\rho(\beta) &= \bar{\rho}(\beta) + \rho_{\rm osc}(\beta).
\end{align}
We study $\bar{\rho}(\beta)$ in Section~\ref{SECT:PXingForMB} and
$\rho_{\rm osc}(\beta)$ in Section~\ref{SECT:DOSOscillations}.

%%%%%%%%%%%%%%%%%%%%%%%%%%%%%%%%%%%%%
%									%
%	PXing for MB					%
%									%
%%%%%%%%%%%%%%%%%%%%%%%%%%%%%%%%%%%%%

\section{Average density of fermion-parity crossings}\label{SECT:PXingForMB}

%%%%%%%%%%%%%%%%%%%%%%%%%%%%%%%%%%%%%
%									%
%	Figure 2: Weyl DOS				%
%									%
%%%%%%%%%%%%%%%%%%%%%%%%%%%%%%%%%%%%%

\begin{figure}[tb]
\centerline{\includegraphics[width=1\linewidth]{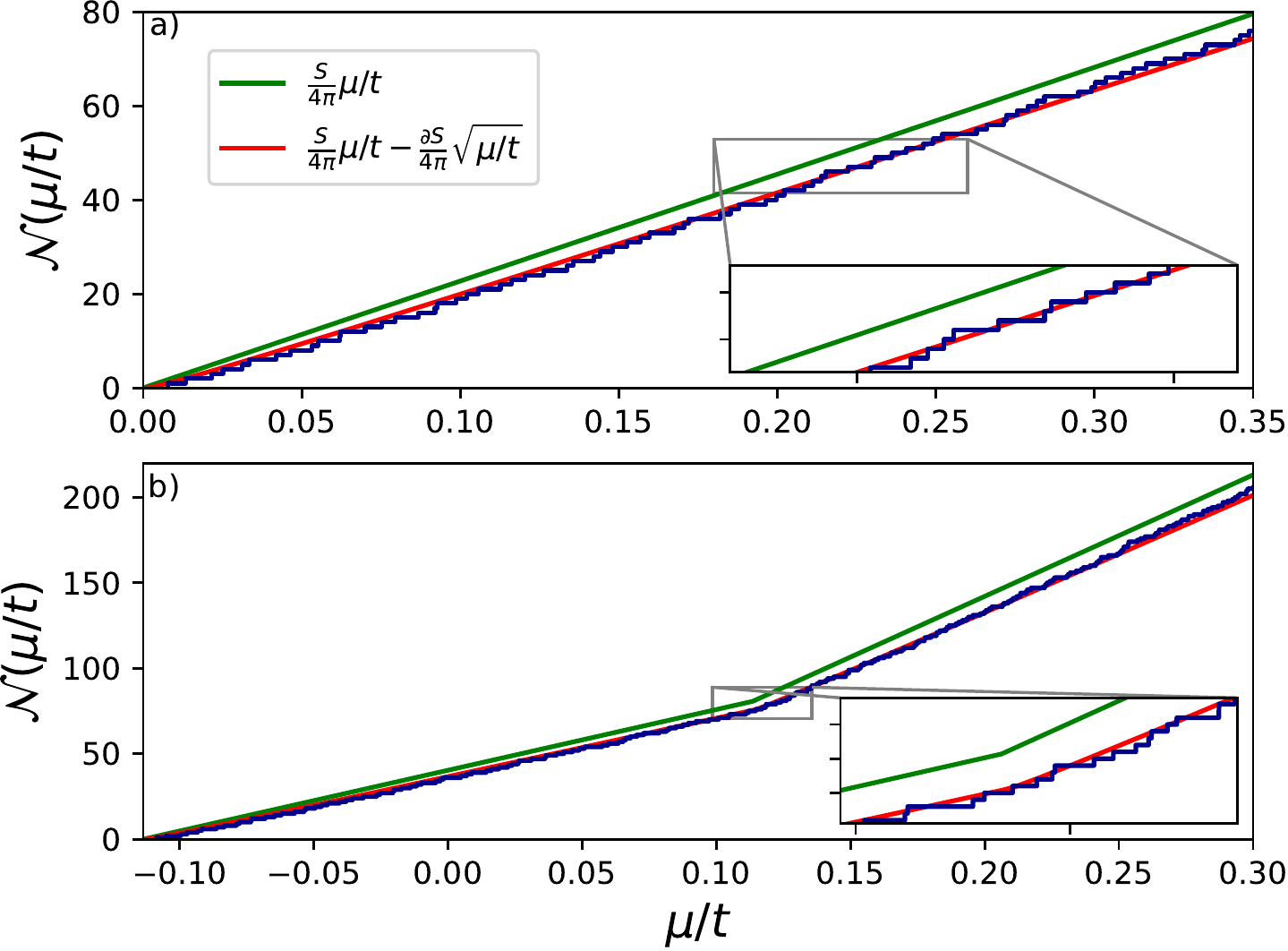}}
\caption{(Color online) $\mathcal{N}(\mu/t)$ for a ballistic quarter
stadium MB (see Fig.~\ref{FIG:2DSystemShape}c. The solid lines are
obtained using Eq.~(\ref{EQN:WeylExpansion_PWave}) for the top panel
and Eq.~(\ref{EQN:WeylExpansion_SWave}) for the bottom panel, as a
function of $\mu/t$.  The green line refers to the first term in the
Weyl expansion whereas the red line includes the surface corrections.
The staircase plot (blue line) is the result of tight-binding
simulations. Lower-right insets are zoom-ins to show the fit between
tight-binding simulation and theory. a) \textit{p}-wave Majorana
billiard with with $L=80a$, $W=40a$ and $\Delta' = 0.001 ta$. b)
\textit{s}-wave MB with $L=100a$, $W=50a$, $B = 0.23t$, $\Delta=0.2t$
and $\alpha = 0.001 ta$. The kink in the plot is at $\mu=\epsilon$ and
signals the entrance of the second spin band, previously
spin-polarized,  into the picture.} \label{FIG:2DWeyl}
\end{figure}

In this section, we investigate the density of FPXs for \textit{p}- and
\textit{s}-wave topological superconductors. We show that the FPX
points of $H_p$ and $H_s$ are real eigenvalues of a corresponding
non-Hermitian operator (Eqs.~(\ref{EQN:ModifiedHamiltonian_PWave}) and
(\ref{EQN:ModifiedHamiltonian_SWave})). Further simplification is
possible if $S \ll \xi \partial S$, where $S$ is the system area,
$\partial S$ is the size of the boundary and $\xi$ is the
superconducting coherence length. (For example, for a rectangular
cavity of width $W$, this limit corresponds to $W \ll \xi$ .) In this
limit, the non-Hermitian eigenvalue problem for $H_p$ and $H_s$ can be
transformed by a local rescaling transformation to a Hermitian
eigenvalue problem (Eqs.~(\ref{EQN:NormalStateHamiltonian_PWave}) and
(\ref{EQN:NormalStateHamiltonian_SWave})). We thus show that,
surprisingly, the FPX points of MBs are related to the energy
eigenvalues of a Hermitian operator which we identify as the normal
state Hamiltonian. We next derive the Weyl expansion for the average
density of FPXs, which is expressed in
Eqs.~(\ref{EQN:WeylExpansion_PWave}) and
(\ref{EQN:WeylExpansion_SWave}) for the \textit{p}- and \textit{s}-wave
cases, respectively. We also perform numerical tight-binding
simulations, which we detail in
Appendix~\ref{SECT:Appendix:TB_Methods}, and compare our results with
our formulae. We present our results for a 2D Majorana billiard in
Figs.~\ref{FIG:2DWeyl}a and b, where we plot the integrated density
of FPXs $\mathcal{N}(\mu/t)$ for \textit{p}- and \textit{s}-wave
systems.  We see that the analytical and numerical results fit
remarkably well without any fitting parameters, once the boundary
corrections in the Weyl expansion are taken into account.

%%%%%%%%%%%%%%%%%%%%%%%%%%%%%%%%%%%%%
%									%
%	Map to Weyl Problem - p wave	%
%									%
%%%%%%%%%%%%%%%%%%%%%%%%%%%%%%%%%%%%%

\subsection{Average density of FPXs of a \textit{p}-wave Majorana billiard}\label{SUBSECT:MapToWeylPWave}

We first focus on the FPXs of a \textit{p}-wave Majorana billiard
described by the Hamiltonian $H_p$ (Eq.~(\ref{EQN:Hamiltonian_PWave})).
In this case, there's only a single external parameter, namely the
chemical potential, to be varied, hence $\beta = \mu/t$. The FPX points
are the $\mu_i$ values for which the \textit{p}-wave h Hamiltonian has
a zero-energy eigenstate:
\begin{align}\label{EQN:PWave_PxingPointZeroEnergy}
H_p|_{\mu=\mu_i} \, \chi &= 0.
\end{align}
We map the problem of finding the FPX points to that of finding
eigenvalues of a non-Hermitian operator by premultiplying
Eq.~(\ref{EQN:PWave_PxingPointZeroEnergy}) by $\tau_z$:
\begin{align}\label{EQN:ModifiedHamiltonian_PWave}
\Bigg(\frac{\big(\mathbf{p}+ i m \Delta' \boldsymbol{\eta}\big)^2}{2m} + V(\mathbf{r}) +m\Delta'^2
\Bigg)\chi &= \mu\, \chi,
\end{align}
where $\boldsymbol{\eta} = \tau_y \hat{x} - \tau_x \hat{y}$. We
identify this operator as the Hamiltonian of a Rashba 2DEG with an
imaginary Rashba parameter $\alpha=i\Delta'$.
Eq.~(\ref{EQN:ModifiedHamiltonian_PWave}) shows that the real
right-eigenvalues of this non-Hermitean operator correspond to the FPX
points, whereas the complex eigenvalues are associated with avoided
crossings.

There is no general reason to assume that a given right-eigenvalue of
Eq.~(\ref{EQN:ModifiedHamiltonian_PWave}) is real. However,
further simplification is possible in the limit of $S/\partial S \ll
\xi=\hbar/m\Delta'$. Rescaling the eigenfunction $\chi= {\rm
e}^{\boldsymbol{\eta} \cdot \mathbf{r}/\xi-r^2/\xi^2}\tilde{\chi}$ and
expanding in powers of $S/(\xi \, \partial S)$, we
obtain~\cite{REF:Adagideli14}
\begin{align}\label{EQN:PWave_MuEigenvalue}
\Bigg( \frac{\big(\mathbf{p}+\frac{2m^2\Delta'^2}{\hbar} (\hat{\mathbf{z}}\times \mathbf{r})\,\tau_z \big)^2}{2m}
	+ V(\mathbf{r}) + m\Delta'^{2} \Bigg)\, \tilde{\chi}
	&=\mu \, \tilde{\chi}.
\end{align}
We see that the crossing points are eigenvalues of the normal state
Hamiltonian with a fictitious magnetic field $\pm
2m^2(\Delta')^2/e\hbar$ and a constant potential shift $m(\Delta')^2$.
We further note that the energy levels are even functions of applied
magnetic fields. Therefore, to the order we are working in, the effect
of the fictitious magnetic field on the crossing points can be ignored,
as they only serve to modify the nonzero split in energy levels. Hence
we see that all eigenvalues of Eq.~(\ref{EQN:PWave_MuEigenvalue})
are real. We thus arrive at the remarkable result that all FPX points
are simply eigenvalues of a \emph{normal state} Hamiltonian:
\begin{align}\label{EQN:NormalStateHamiltonian_PWave}
\left(\frac{p^2}{2m} + V(\mathbf{r}) + m\Delta'^2 \right)\,\tilde{\chi} &= \mu \tilde{\chi}.
\end{align}
This identification allows us to map the average density of FPXs to the
conventional density of states of a normal state Hamiltonian. Well
known results, such as the Weyl expansion for average
DOS~\cite{REF:Book:Weyl68, REF:Balian70, REF:Book:Baltes76} (or, for
the case of soft confinement, the Thomas-Fermi
approximation~\cite{REF:Book:Brack03}); Gutzwiller's trace formula in
billiards for oscillations (supershell effects) in
DOS~\cite{REF:Book:Gutzwiller90, REF:Jalabert90, REF:Ishio95,
REF:Adagideli02a, REF:Adagideli02}; the theory of Lifshitz
tails~\cite{REF:Lifshitz64, REF:Halperin65, REF:BOOK:Itzykson89} for
disordered systems; as well as the random matrix theory results for DOS
fluctuations~\cite{REF:Beenakker97, REF:Beenakker13b}, carry over to
the spectra of fermion-parity crossings.

For the average density of FPXs for the \textit{p}-wave system 
$\bar{\rho}_{\textrm{w}, p}(\mu)$ in $d$ dimensions, we thus 
obtain~\cite{REF:Footnote:3DDoFPX}:
\begin{align}\label{EQN:WeylExpansion_PWave}
\bar{\rho}_{\textrm{w}, p}(\mu)	&=	\begin{cases}
								\frac{L}{2\pi\sqrt{\mu}}+\mathcal{O}(1) & \textrm{if } d=1 \\
								\frac{S}{4\pi}-\frac{\partial S}{8\pi \sqrt{\mu}}  & \textrm{if } d=2\\
								 \frac{V \sqrt{\mu}}{4\pi^2}-\frac{\partial V}{16\pi} & \textrm{if } d=3,
								\end{cases}
\end{align}
where $L$ is the length of the 1D wire, $S$ and $\partial S$ are the
area and perimeter of the 2D billiard, and $V$ and $\partial V$ the
volume and surface area of the 3D dot cavity respectively.

%%%%%%%%%%%%%%%%%%%%%%%%%%%%%%%%%%%%%
%									%
%	Map to Weyl Problem - s wave	%
%									%
%%%%%%%%%%%%%%%%%%%%%%%%%%%%%%%%%%%%%

\subsection{Average density of FPXs of a \textit{s}-wave Majorana billiard}\label{SUBSECT:MapToWeylSWave}

We now focus on the FPXs of an \textit{s}-wave Majorana billiard
described by $H_s$ (Eq.~(\ref{EQN:Hamiltonian_SWave})). In this case,
there are two external parameters, namely the chemical potential and
the Zeeman energy. Hence $\beta$ can be either $\mu/t$ or $B/t$. We
again start with the zero energy eigenvalue problem

\begin{align}\label{EQN:PWave_SxingPointZeroEnergy}
H_s|_{\mu_i, B_j}\, \psi &= 0
\end{align}
where $\mu_i$ and $B_j$ are the FPX points. Here, we have two
equivalent choices of obtaining a non-Hermitian eigenvalue problem:
eigenvalues corresponding to $B$ or to $\mu$. This equivalence leads to
a scaling relation between $\mu_i$ and $B_j$ which we discuss in
Section~\ref{SUBSECT:SWaveUniversal}. Without loss of generality we
focus on the eigenvalue problem for $\mu_i$ below. We premultiply
Eq.~(\ref{EQN:PWave_SxingPointZeroEnergy}) with $\tau_z$ and obtain
\begin{align}\label{EQN:ModifiedHamiltonian_SWave}
\left( \frac{\mathbf{p}^2}{2m} + V(\mathbf{r}) + \alpha \boldsymbol{\eta}\cdot \mathbf{p} + B \sigma_x \tau_z + i\Delta \tau_y \right) \psi &= \mu \,\psi,
\end{align}
where $\boldsymbol{\eta} = (\sigma_y \hat{x} - \sigma_x \hat{y})$. This
equation can then be solved using tight binding methods, see
appendix~\ref{SECT:Appendix:TB_Methods}.

In order to proceed analytically, we follow
Ref.~[\onlinecite{REF:Adagideli14}] and [\onlinecite{REF:Pekerten17}]
to again transform the usual eigenvalue problem ($H_s \, \psi =
E\,\psi$ with $E=0$) to a non-Hermitian eigenvalue problem and obtain:
\begin{align}\label{EQN:SWave_Rotated}
\left(h(\mathbf{p}, \mathbf{r}) \sigma_z - i\alpha p_x  \sigma_x \mp B \mp \Delta \sigma_x \right)\, \phi_\pm &= 0.
\end{align}
Here, we have ignored the chiral symmetry breaking term $i \alpha p_y
\sigma_y$, which is justified in the limit $S \ll \xi \partial S$, as
in the previous section. For a finite system, the solution that
satisfies all boundary conditions can be expressed as
\begin{equation}\label{EQN:RescaledWavefunction}
\phi_{n,\pm} = \zeta_\pm(E_n) {\rm e}^{\pm x/\xi}  \psi_n,
\end{equation}
where $\zeta_\pm(\epsilon)$ are the eigenvectors  of the $2\times 2$
matrix $\epsilon \, \sigma_z \mp \Delta \sigma_x$ with eigenvalue $\pm
\sqrt{\epsilon^2 +\Delta^2}$ and $ \psi_n$ satisfies the eigenvalue
equation:
\begin{align}\label{EQN:NormalStateHamiltonian_SWave}
h \, \psi_n &= E_n\, \psi_n.
\end{align}

Substituting Eq.~(\ref{EQN:RescaledWavefunction}) into
Eq.~(\ref{EQN:SWave_Rotated}), we find that the zero mode solutions
(hence the fermion-parity crossings) happen on families of curves in
the $B-\mu$ plane. The curves satisfy
\begin{equation}\label{EQN:SWaveXingPoints_B_mu}
B^2=(\mu - E_n)^2 + \Delta^2
\end{equation}
for a given eigenvalue $E_n$ of the spinless single particle
Hamiltonian $h(\mathbf{p}, \mathbf{r})$. Hence, the density of FPX
spectrum (with respect to either $\mu$ or $B$) can be obtained by
analyzing the set of eigenvalues $\{E_n \}$ of $h(\mathbf{p},
\mathbf{r})$. Noting that $h(\mathbf{p}, \mathbf{r})$ is the same for
\textit{s}- and \textit{p}-wave cases, we write the \textit {s}-wave
Weyl expansion for $\rho_{\textrm{w}, s}(\mu)$ and $\rho_{\textrm{w},
s}(B)$ for fermion-parity crossing densities in terms of their \textit
{p}-wave counterpart $\rho_{\textrm{w}, p}(\mu)$ in
Eq.~(\ref{EQN:WeylExpansion_PWave}):
\begin{equation}\label{EQN:WeylExpansion_SWave}
\rho_{\textrm{w}, s}(\mu, B) = \sum_{\varsigma=\pm 1}\rho_{\textrm{w}, p}(\mu+\varsigma\epsilon) \, \theta(\mu + \varsigma\epsilon),
\end{equation}
where $\theta(x)$ is the Heaviside step function,
$\epsilon=\sqrt{B^2-\Delta^2}$ as before and the $\varsigma=\pm 1$
terms in the sum correspond to the densities of different spin species
separated in energy by the Zeeman field.

%%%%%%%%%%%%%%%%%%%%%%%%%%%%%%%%%%%%%%%%%%%%%%%%%
%												%
% Universal SWave fermion-parity Xing Spectra	%
%												%
%%%%%%%%%%%%%%%%%%%%%%%%%%%%%%%%%%%%%%%%%%%%%%%%%

\subsection{Universal scaling properties of fermion-parity crossing
points in \textit{s}-wave systems}\label{SUBSECT:SWaveUniversal}

%%%%%%%%%%%%%%%%%%%%%%%%%%%%%%%%%%%%%
%									%
%	Figure: SWave Pxing points 		%
%									%
%%%%%%%%%%%%%%%%%%%%%%%%%%%%%%%%%%%%%

\begin{figure}
\includegraphics[width=\linewidth]{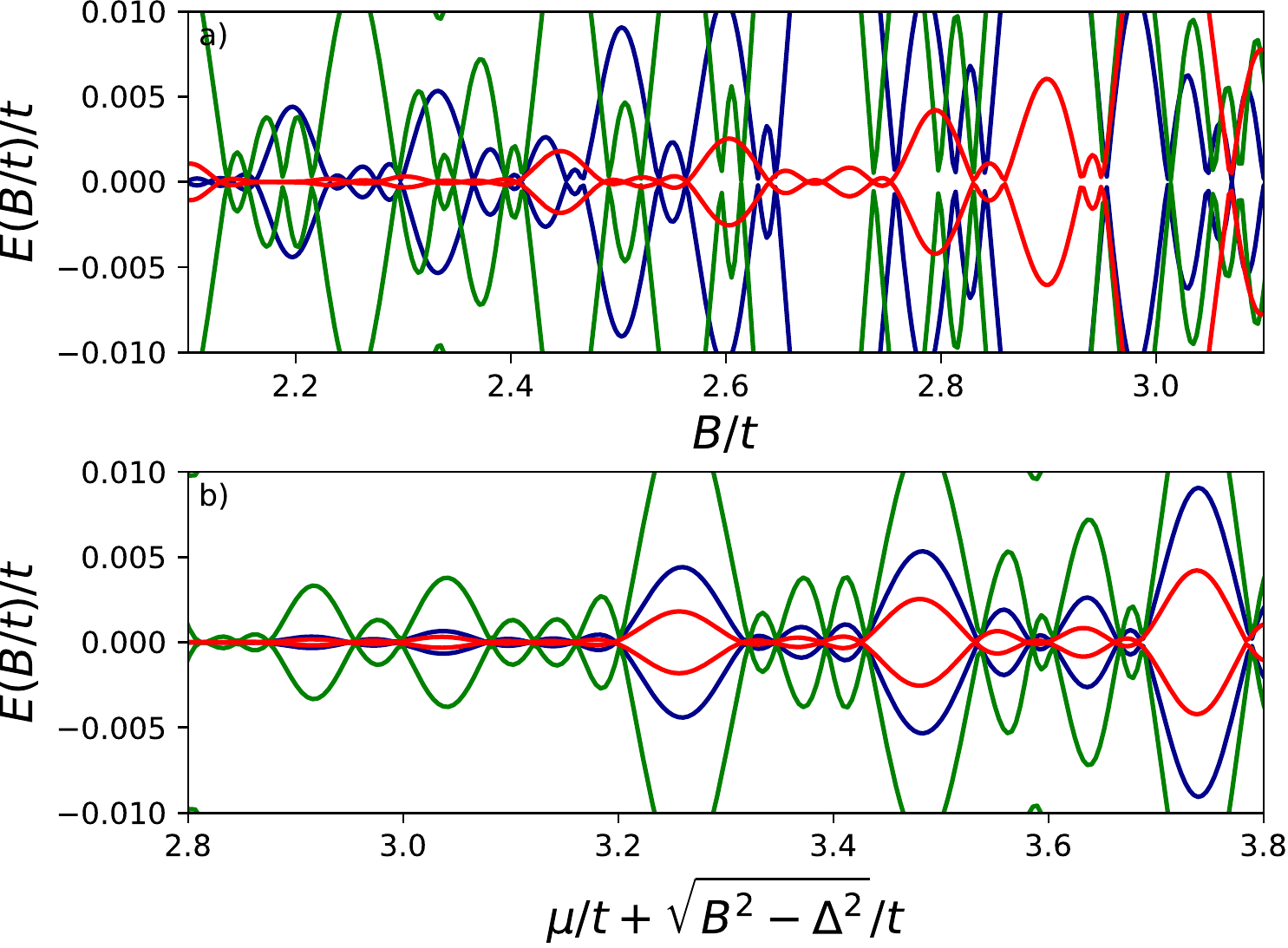}
\caption{(Color online) A plot of the lowest four eigenvalues of the
disordered \textit{s}-wave Hamiltonian in
Eq.~(\ref{EQN:Hamiltonian_SWave}), discretized on a 1D lattice of 100
sites, plotted as a function of (a) $B/t$ and (b) $\mu/t +
\sqrt{B^2-\Delta^2}/t$, for different values of Hamiltonian parameters.
In both plots, the green set of curves represents the lowest four
eigenvalues obtained for $\Delta=1.5t$, $\alpha = 0.05ta$, $\mu=1.8t$;
the blue set is for $\Delta=1.8t$, $\alpha = 0.05ta$, $\mu=2.0t$; and
the red set is for $\Delta=1.8t$, $\alpha = 0.08ta$, $\mu=1.6t$. In all
cases, the same disorder realization with a disorder strength $V_d =
0.5t$ is utilized.}\label{FIG:SWaveXing_Shifted}
\end{figure}

As a consequence of Eq.~(\ref{EQN:SWaveXingPoints_B_mu}), the FPX
spectra exhibit a scaling relation for a given disorder realization:
all the FPXs corresponding to different values of $\mu$, $B$ or
$\Delta$, collapse on the same set of points if expressed in terms of
the combination $\mu \pm \sqrt{B^2-\Delta^2}$
(Fig.~\ref{FIG:SWaveXing_Shifted}). Moreover, if the FPX spectrum of
one of the Zeeman-split spin bands is known, the other can immediately
be determined by shifting the spectrum by $2 \sqrt{B^2-\Delta^2}$.

This universality is evident in Fig.~\ref{FIG:SWaveXing_Shifted}, where
we plot the first four eigenvalues of a 1D \textit{s}-wave system with
a specific disorder realization for different values of $\mu$ and
$\Delta$ as a function of $B$ in Fig.~\ref{FIG:SWaveXing_Shifted}a
and as a function of $\mu+\sqrt{B^2-\Delta^2}$ in
Fig.~\ref{FIG:SWaveXing_Shifted}b. These plots are obtained by
discretizing the \textit{s}-wave Hamiltonian in
Eq.~(\ref{EQN:Hamiltonian_SWave}) in 1D over $100$ sites and
numerically solving the resulting eigenvalue problem. We see that in
Fig.~\ref{FIG:SWaveXing_Shifted}b, all energy level crossings happen
at the same set of values of $\mu+ \sqrt{B^2-\Delta^2}$ for systems
with the same disorder realization but different system parameters.

%%%%%%%%%%%%%%%%%%%%%%%%%%%%%%%%%%%%%
%									%
%	Lifshitz tail					%
%									%
%%%%%%%%%%%%%%%%%%%%%%%%%%%%%%%%%%%%%

\subsection{Lifshitz tail in disordered MBs}\label{SECT:Appendix:LifshitzTail}

%%%%%%%%%%%%%%%%%%%%%%%%%%%%%%%%%%%%%
%									%
%	Figure: Lifshitz tail			%
%									%
%%%%%%%%%%%%%%%%%%%%%%%%%%%%%%%%%%%%%

\begin{figure}
\includegraphics[width=\linewidth]{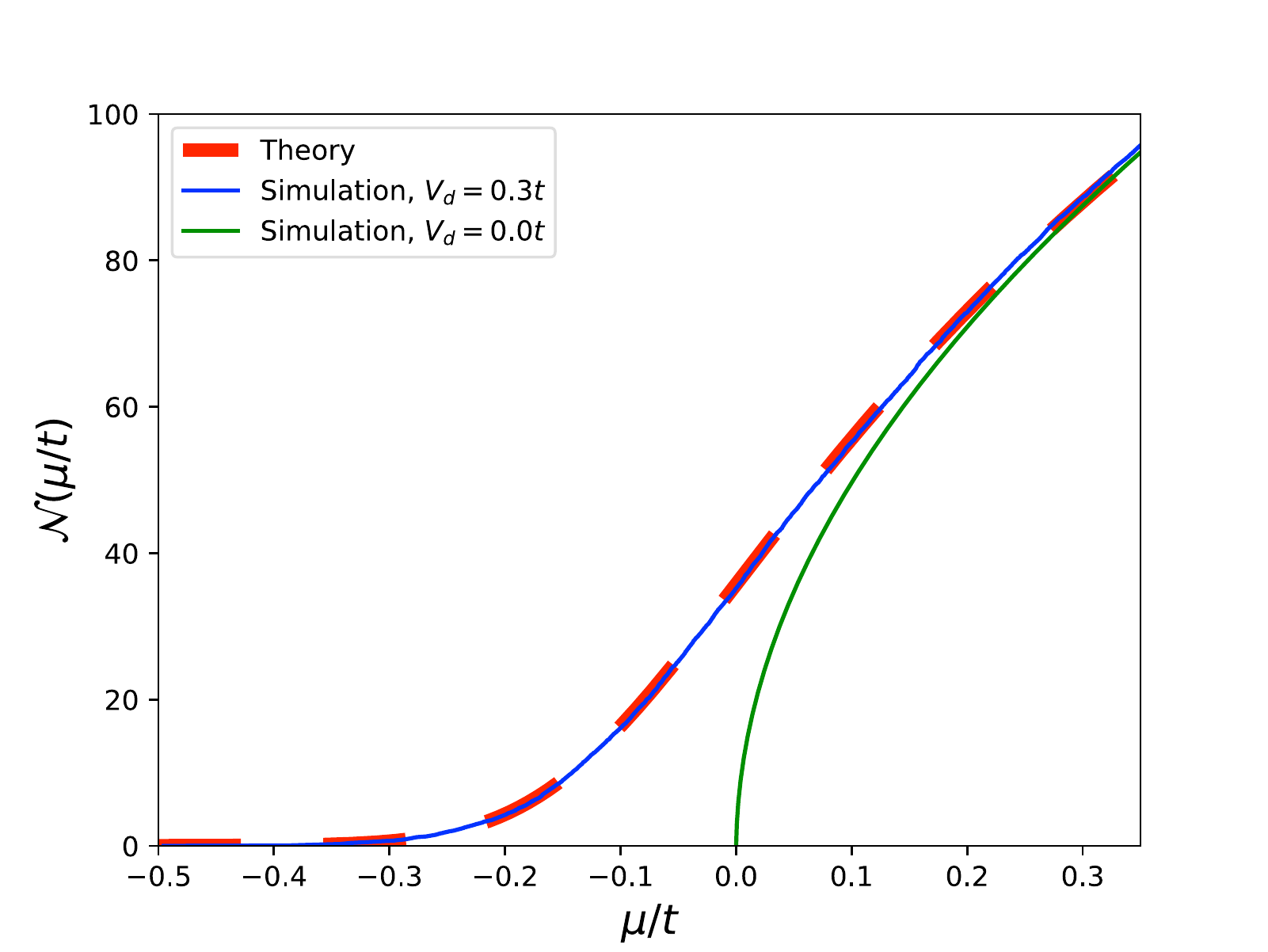}
\caption{(Color online) $\mathcal{N}(\mu/t)$ vs. $\mu/t$ for a
\textit{p}-wave 1D MB for a wire of length $500a$ and $\Delta'=0.001
ta$. For the disordered case, the tight-binding simulation plot is the
average of $200$ disorder realizations. The theory lines are the plots
of Eq.~(\ref{EQN:Appendix:N_E_Airy}) for $V_d=0$ and $V_d=0.3t$.}\label{FIG:Appendix:PWMultipleDisorder}
\end{figure}

Disordered systems feature states below zero energy due to the presence
of islands with an average of below zero potential, even though the
average potential for the whole system is zero. Called the Lifshitz
tail~\cite{REF:Lifshitz64, REF:Halperin65, REF:BOOK:Itzykson89}, this
phenomenon is also present in density of FPXs in MBs (see
Fig.~\ref{FIG:Appendix:PWMultipleDisorder}). The overall
disorder-averaged integrated density of FPXs
$\mathcal{N}(\mu/t)$ for a 1D \textit{p}-wave MB with Gaussian disorder
(i.e. $\left\langle V(\mathbf{r}) V(\mathbf{r}') \right\rangle =D \,
\delta(\mathbf{r}-\mathbf{r}')$ ) is given by the
formula~\cite{REF:BOOK:Itzykson89}:
\begin{align}\label{EQN:Appendix:N_E_Airy}
\mathcal{N}(\mu)	&=	\frac{\kappa_0}{\pi^2\, \varepsilon_0} \,
									\frac{1}{[\textrm{Ai}(-2\mu/\varepsilon_0)]^2 +
									[\textrm{Bi}(-2\mu/\varepsilon_0)]^2},
\end{align}
where $\textrm{Ai}$ and $\textrm{Bi}$ are the Airy functions,
$\varepsilon_0 = (D^2 \, m \hbar^{-2})^{1/3}$ and $\kappa_0 = (D \, m^2
\hbar^{-4})^{1/3}$.

In Fig.~\ref{FIG:Appendix:PWMultipleDisorder}, we plot
Eq.~(\ref{EQN:Appendix:N_E_Airy}) and tight-binding simulations
for a 1D disordered wire (and a tight-binding simulation for the same
wire with zero disorder for comparison). We observe FPXs in the fully
spin-polarized wire even in negative values of $\mu$, caused by rare
disorder configurations. We note that the theory and the numerical
simulations show remarkable agreement without any fitting parameters.

%%%%%%%%%%%%%%%%%%%%%%%%%%%%%%%%%%%%%
%									%
%	DOS OScillations				%
%									%
%%%%%%%%%%%%%%%%%%%%%%%%%%%%%%%%%%%%%

\section{Oscillatory part of density of fermion-parity crossings
}\label{SECT:DOSOscillations}

%%%%%%%%%%%%%%%%%%%%%%%%%%%%%%%%%%%%%
%									%
%	Figure 3: DOS Oscillations		%
%									%
%%%%%%%%%%%%%%%%%%%%%%%%%%%%%%%%%%%%%

\begin{figure}[tb]
\centerline{\includegraphics[width=1\linewidth]{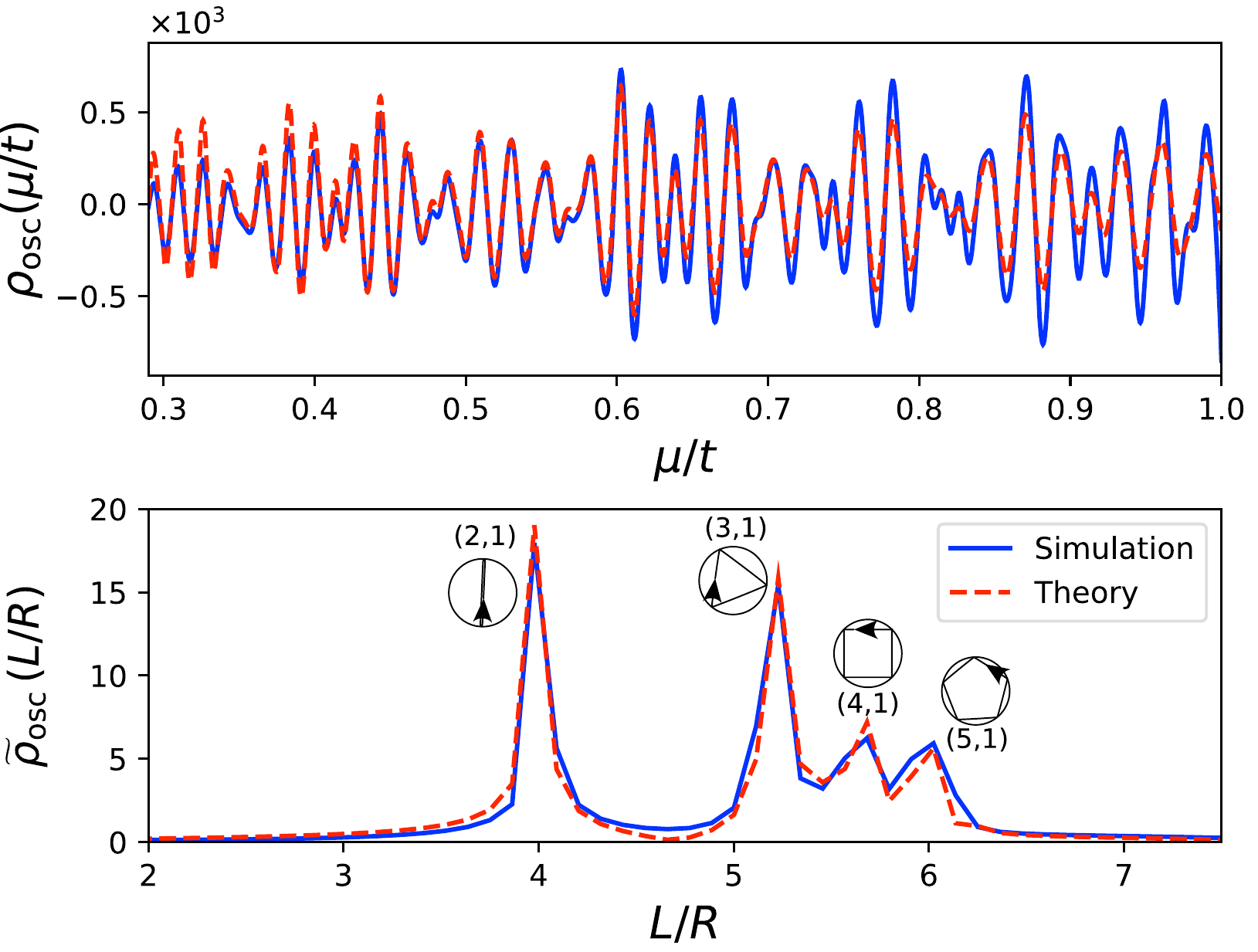}}
\caption{(Color online) a) Density oscillations of fermion-parity
crossings $\rho_\textrm{osc}$ for a clean \textit{p}-wave disk Majorana
billiard with $R = 100 a$, $\Delta' = 0.001 ta$. b) The Fourier
transform of $\rho_{osc}$. The $(v, w)$ pairs and corresponding
classical orbits for the peaks are labeled. The smoothing parameter for
both figures is $\gamma = 0.4/R$.}\label{FIG:DensityOscillations}
\end{figure}

We next investigate the oscillatory part $\rho_{\rm osc}$ of the
density of FPXs (see Eq.~(\ref{EQN:DOSMeanPlusOsc})). The DOS analog of
such oscillations are the so-called shell and supershell effects known
from the studies of finite quantum systems such as nuclei, atomic
clusters and nanoparticles. The celebrated Guztwiller or Balian-Bloch
trace formula show that each periodic orbit contributes a term
oscillating with its classical action~\cite{REF:Balian70,
REF:Book:Gutzwiller90, REF:Jalabert90, REF:Ishio95, REF:Adagideli02a,
REF:Adagideli02}.

In this section, we extend the analysis of the oscillatory part of DOS
in Ref.~[\onlinecite{REF:Balian70}] and [\onlinecite{REF:Book:Brack03}]
to the case of the FPX spectrum of a clean \textit{p}-wave MB. We again
take advantage of the mapping described in
Section~\ref{SUBSECT:MapToWeylPWave} of the \textit{p}-wave Hamiltonian
to a normal state Hamiltonian with eigenvalues yielding the FPX points.
We thus extend the Gutzwiller and/or Balian Bloch trace
formula~\cite{REF:Balian70,REF:Book:Gutzwiller90} from its original
setting of the DOS of finite systems into the FPXs of finite Majorana
platforms. The new trace formula expresses the oscillating part
$\rho_\textrm{osc}$ as a sum over classical periodic orbits $\zeta$.
Its general form is
\begin{equation}
\label{EQN:2DTraceFormula_Degenerate}
\rho_\textrm{osc}(\mu)= \sum_\zeta{\cal A}_\zeta\cos \Phi_\zeta(\mu),
\end{equation}
where ${\cal A}_\zeta$ is related to the stability of the orbit and
$\hbar \Phi_\zeta$ is related to its classical action as well as the
Maslov indices. Their detailed form depends on whether the orbits are
isolated or part of a family of orbits (sometimes called degenerate
orbits). For isolated periodic orbits,
\begin{align}
{\cal A}_\zeta = \frac{T_{\zeta}/\pi\hbar}{\sqrt{\vert\det(M_\zeta-I) \vert}} ,\quad
\Phi_\zeta(\mu)= \frac{S_\zeta(\mu)}{\hbar}- \frac{\sigma_\gamma\pi}{2},
\end{align}
where $T_{\zeta}$ is the period of the corresponding primitive periodic
orbit (i.e. the parent orbit with no retracings), $M_\zeta$ is the
stability matrix of the orbit~\cite{REF:FOOTNOTE:stabilityM} and
$\sigma_\gamma$ is the Maslov index. The final ingredient is the
classical action, given by $S_\zeta(\mu)= \oint_\zeta \mathbf{p} \cdot
d\mathbf{r}$. The weight of individual contributions increases for
degenerate orbits. For two dimensional systems--which is our main
focus--and singly degenerate orbits
\begin{align}
&{\cal A}_\zeta  = \frac{2m}{(2\pi \hbar)^{3/2} p_F}
					\int \left|\frac{\partial r_\perp}{\partial p'_\perp}\right|^{-1/2}_\zeta dr_\parallel\, dr_\perp \, , \nonumber \\
&\Phi_\zeta(\mu) = \frac{S_\zeta(\mu)}{\hbar}- \frac{\sigma_\gamma\pi}{2}- \frac{\pi}{4},
\end{align}
where $p_F$ is the Fermi momentum. Here an initial transverse
perturbation of momentum $p_\perp'$ leads to a final transverse
deviation $r_\perp$ after a full round. We note that in a billiard
system $\vert \mathbf{p}\vert =p_F$, hence the classical action
corresponding to a periodic orbit is $S_\zeta(\mu)=p_F L_\zeta $ where
$L_\zeta$ is the length of the orbit $\zeta $.

In order to demonstrate our results, we specialize to a clean
\textit{p}-wave disk MB of radius $R$ (see
Fig.~\ref{FIG:2DSystemShape}). For this system, it is possible to
obtain closed-form analytical formulae using
Eq.~(\ref{EQN:2DTraceFormula_Degenerate}) and compare the numerical
simulations with these formulae. We first note that a periodic orbit of
a disk billiard is uniquely determined by the number $w$ times the
orbit winds around the billiard and the number $v$ times it reflects
from the boundary. Then a simple geometrical consideration allows one
to express the length of  the orbit as $L_{vw} =2vR\sin(\pi w/v)$. We
thus obtain
\begin{align}\label{EQN:rhoOsc_PWaveDisk_General}
\rho_\mathrm{osc}(\mu) 	&= 	\frac{2 m R^2}{\hbar^2}\, \bigg(\frac{\hbar}{\pi R \, p(\mu)} \bigg)^{1/2}\, \nonumber\\
						&\quad \times \sum_{w=1}^{\infty} \sum_{v=2w}^{\infty} f_{vw}\frac{\sin^{3/2}(\pi w/v	)}{\sqrt{v}} \nonumber\\
						&\quad \times \mathrm{Im}\bigg[\exp\bigg\{i \frac{p_FL_{vw}}{\hbar} +i\phi_\textrm{po} \bigg\}\bigg],
\end{align}
where $\phi_\textrm{po} = - 3v\pi/2 + 3\pi/4$, $f_{vw} = 2 \,
\theta(v-2w)$ with $\theta(x)$ being the Heaviside step function. In
Fig.~\ref{FIG:DensityOscillations}a, we plot $\rho_\textrm{osc}(\mu/t)$
as determined from numerical solutions of the Majorana
billiard~\cite{REF:Scharf2015, REF:FOOTNOTE:Scharf} (blue, solid line)
and as given by Eq.~(\ref{EQN:rhoOsc_PWaveDisk_General}) (red, dashed
line) for a \textit{p}-wave disk MB. Both lines are smoothed using a
Gaussian smoothing function. The plots show remarkable agreement. In
Fig.~\ref{FIG:DensityOscillations}b, we plot the Fourier transform
$\tilde{\rho}_\textrm{osc} (L/R)$ of
Fig.~\ref{FIG:DensityOscillations}a in order to observe the location of
the periodic orbits and their relative amplitudes. (We choose to show
the Fourier transform as a function of the dimensionless parameter
$L/R$, i.e. orbit length divided by disk radius, rather than as a
function of the period of the orbit for convenience, since the length
and the period of a given orbit are proportional.)  As discussed above,
the peaks are centered around the $L/R$ values of the high-degeneracy
orbits (shown in the insets) and their relative amplitude reflects
their order of degeneracy.

It is a straightforward task to extend
Eq.~(\ref{EQN:rhoOsc_PWaveDisk_General}) for the case of a generic
(tight-binding) energy dispersion and obtain the corresponding
$\rho_\textrm{osc}$, for details we refer the reader to
Appendix~\ref{SECT:Appendix:Oscillations}.

%%%%%%%%%%%%%%%%%%%%%%%%%%%%%%%%%%%%%
%									%
%	Universal statistics			%
%									%
%%%%%%%%%%%%%%%%%%%%%%%%%%%%%%%%%%%%%

\section{Universal fluctuations of fermion-parity crossings}\label{SECT:UniversalStats}

We now focus on how consecutive fermion-parity crossings are
correlated. We first work in the limit $S/\partial S \ll \xi$ (i.e.~one
of the system size parameters (the ``width'') becomes smaller than the
superconducting coherence length) and we obtain the FPX spacing
distributions. We find that the FPX points are uncorrelated for systems
that are localized in their normal state and the spacing distribution
is Poissonian:
\begin{equation}\label{EQN:Poisson}
P(\delta \mu)= \exp \big(-\delta\mu/\langle \delta\mu \rangle\big),
\end{equation}
where $\delta \mu$ is the FPX spacing and $\langle \delta\mu \rangle$
is its ensemble-averaged value. When the normal state system is near a
delocalization transition, the FPX points become correlated and feature
antibunching for small spacings, while large spacings remain
uncorrelated. This behaviour is reflected in the semi-Poissonian
distribution, signaling the fractal nature of the wavefunction near the
metal insulator transition~\cite{REF:Shklovskii93}:
\begin{equation}\label{EQN:SemiPoisson}
P(\delta \mu)=\frac{\delta\mu}{\langle \delta\mu \rangle} \exp \big(-2\delta\mu/\langle \delta\mu \rangle \big).
\end{equation}
Finally if the normal system is delocalized enough that the escape time
is shorter than $\hbar/\langle \delta\mu \rangle$, the FPX points
feature correlations that are reminiscent of the eigenvalues of an
ensemble of real Hermitian random matrices and the corresponding
distribution is the Wigner-Dyson distribution for orthogonal
matrices~\cite{REF:Wigner55, REF:Dyson62, REF:Dyson62b,
REF:Dyson63,REF:Beenakker97, REF:Book:Mehta04}:
\begin{equation}\label{EQN:WignerDyson}
P(\delta \mu)=\frac{\pi\delta\mu}{2 \langle \delta\mu \rangle} \exp \bigg(-\frac{\pi\delta\mu^2}{4\langle \delta\mu \rangle^2} \bigg),
\end{equation}

%%%%%%%%%%%%%%%%%%%%%%%%%%%%%%%%%%%%%
%									%
%	Figure: PWave Statistics		%
%									%
%%%%%%%%%%%%%%%%%%%%%%%%%%%%%%%%%%%%%

\begin{figure}[tb]
\centerline{\includegraphics[width=1\linewidth]{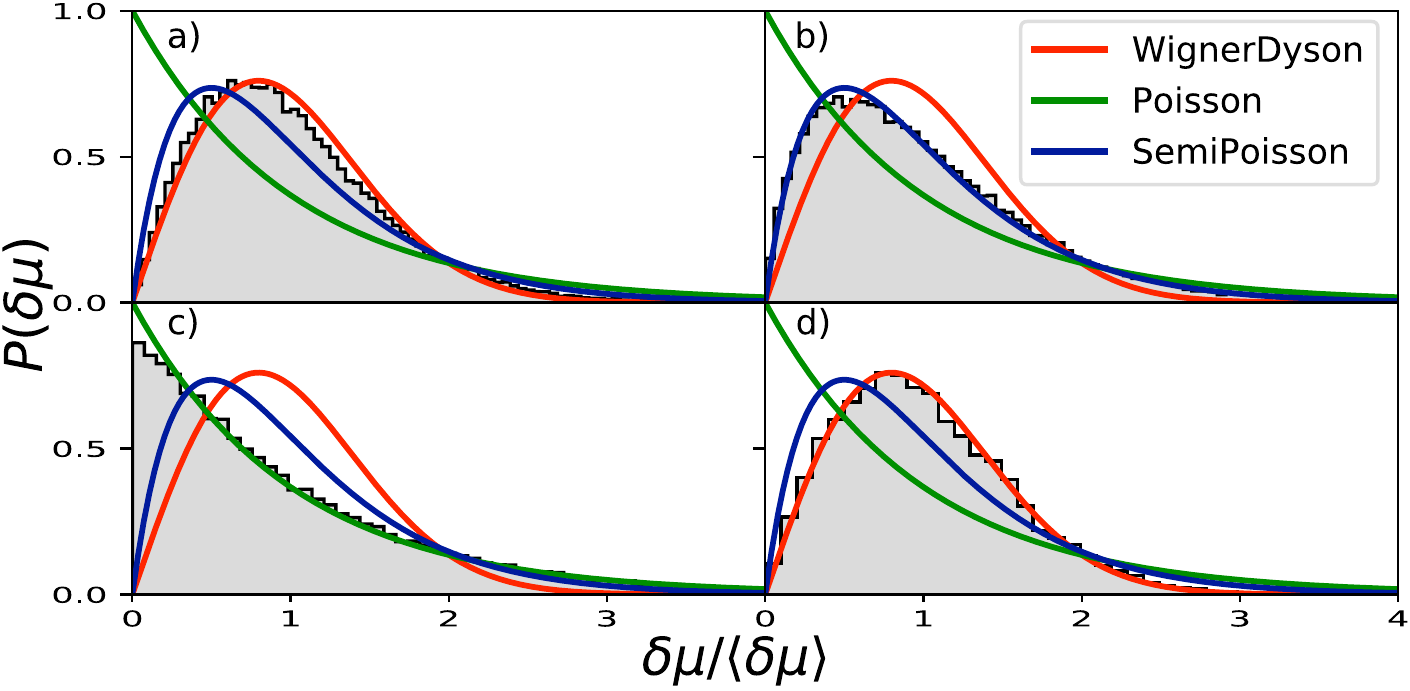}}
\caption{(Color online) a-c) Level spacing distributions for a
disordered rectangular \textit{p}-wave MBs of varying lengths, averaged
over 500 disorder realizations, with $\Delta' = 0.025 ta$, disorder
strength $V_d = 0.5t$, width  $W = 20a$. a) $L = 40a < \xi$, b) $L =
100a\gtrsim \xi$  and c) $L = 1600a \gg \xi$, with $\xi = 80a$ being
the superconducting coherence length. d) Level spacing distributions,
averaged over 225 cavity realizations, for a  clean \textit{p}-wave
Lorentz cavity MB. Here, $\Delta'=0.001 ta$, $L = 50a$, $W= 50a$, and
$r_1=r_2=10a$. The values of $L/\xi$ in panels a)-d) are $0.5$, $1.25$,
$20$ and $0.4$, respectively. }\label{FIG:PWaveStatistics}
\end{figure}

%%%%%%%%%%%%%%%%%%%%%%%%%%%%%%%%%%%%%
%									%
%	Figure: SWave Statistics		%
%									%
%%%%%%%%%%%%%%%%%%%%%%%%%%%%%%%%%%%%%

\begin{figure}[tb]
\centerline{\includegraphics[width=1\linewidth]{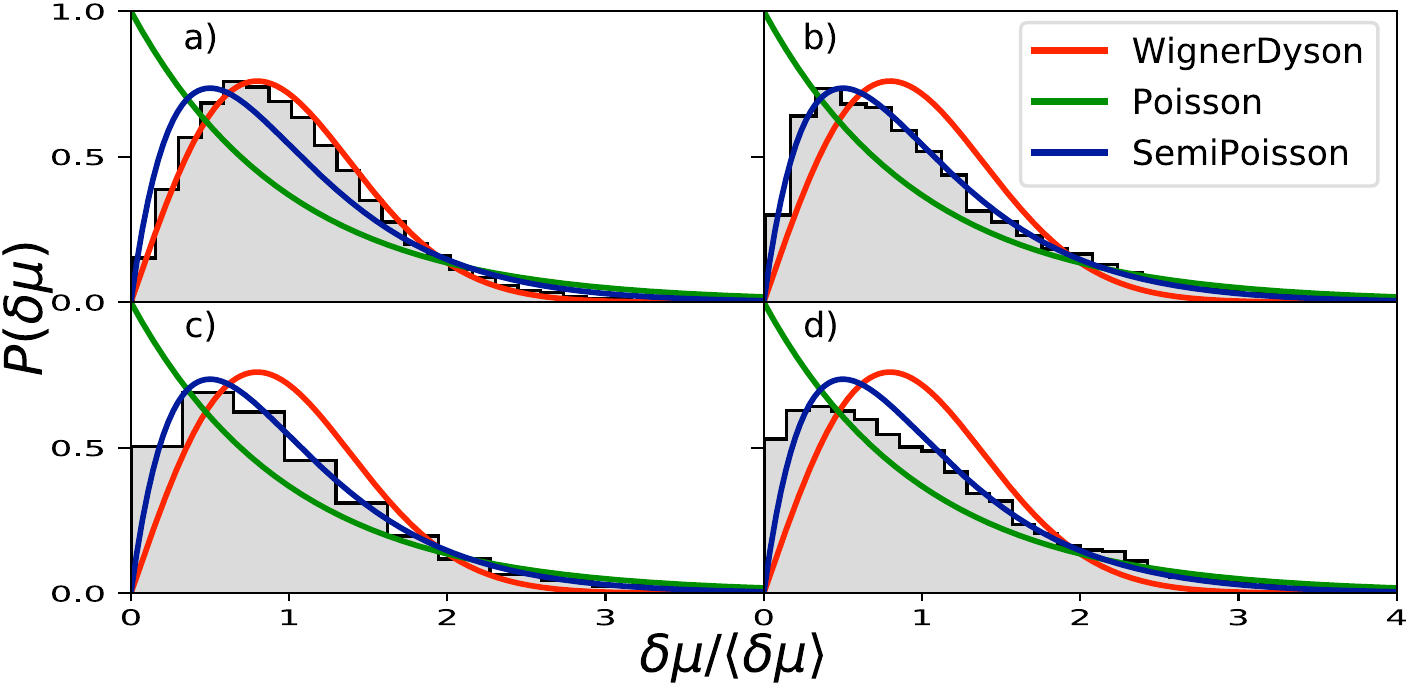}}
\caption{(Color online) a-c) Level spacing distributions for disordered
rectangular \textit{s}-wave MBs with increasing Zeeman energy $B$,
averaged over 500 disorder realizations, with $L=200a$, $W=10a$, $V_d =
0.2t$, $\alpha = 0.025 ta$, $\Delta = 0.12t$, and a) $B = 1.12t$, b) $B
= 0.22t$ and c) $B = 0.13t$. d) Level spacing distributions for clean
\textit{s}-wave Lorentz cavity MB, averaged over 225 cavity
realizations. Here, $\alpha=0.001 ta$, $\Delta = 0.2t$, $B = 0.23t$, $L
= 50a$, $W= 50a$, and $r_1=r_2=10a$. The values of $L/\xi$ in panels
a)-d) are $0.27$, $1.63$, $6.1$ and $0.04$, respectively.}\label{FIG:SWaveStatistics}
\end{figure}

We again utilize a tight-binding model in order to numerically obtain
the FPX spacings and plot the results against the distribution
functions given in Eq.~(\ref{EQN:Poisson}), (\ref{EQN:SemiPoisson}) and
(\ref{EQN:WignerDyson}). Fig.~\ref{FIG:PWaveStatistics}
[Fig.~\ref{FIG:SWaveStatistics}] shows our \textit{p}-wave
[\textit{s}-wave] results for disordered rectangle cavities (a-c) and
chaotic billiards (d). In agreement with our predictions, the
distributions evolve from Wigner-Dyson to semi-Poissonian to Poissonian
as the escape time is increased (the system becomes more localized),
and fit the respective distributions well (see
Fig.~\ref{FIG:PWaveStatistics}). We note, however, that in the
\textit{s}-wave case, $P(\delta \mu \rightarrow 0)$ approaches $0.5$ if
both spin species are populated. This is  due to FPX points
constituting two interlaced sequences belonging to different spin
species~\cite{REF:Pekerten17} for larger $B$ (see
Eq.~(\ref{EQN:WeylExpansion_SWave})). While the elements of each
sequence feature level repulsion, one sequence is the shifted version
of the other. For large enough shifts, the two sequences become
uncorrelated, hence the consecutive spacings between FPX of differing
sequences will also be uncorrelated, suppressing the level repulsion.

\begin{figure}[tb]
\centerline{\includegraphics[width=1\linewidth]{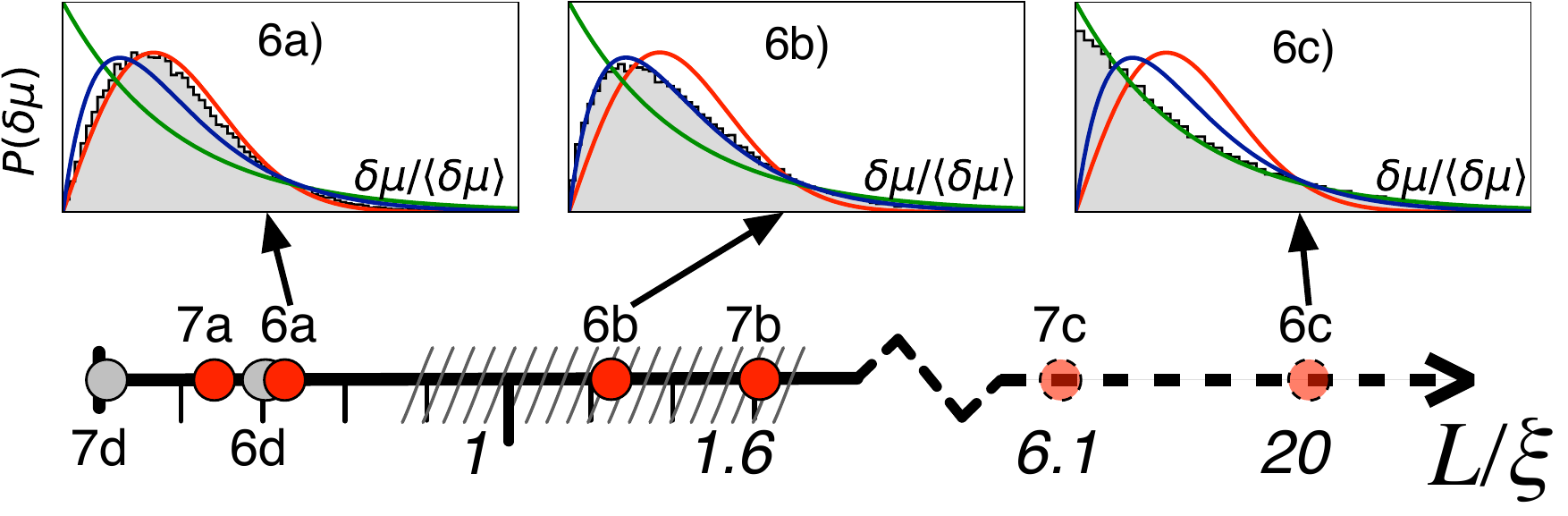}}
\caption{(Color online) The $L/\xi$ values for
Figs.~\ref{FIG:PWaveStatistics}a-d and \ref{FIG:SWaveStatistics}a-d.
The shaded region on the $L/\xi$ axis around $L/\xi = 1$ schematically
represents the universality crossover region where the statistics are
semi-Poissonian. Three panes from Fig.~\ref{FIG:PWaveStatistics} are
reproduced as an example of Gaussian, semi-Poissonian and Poissonian
statistics. Here, $L$ for each shape is defined in
Fig.~\ref{FIG:2DSystemShape}.}\label{FIG:UniversalityCrossoverSmallW}
\end{figure}

Finally, we demonstrate a crossover between the universality classes in
thin ($W \ll \xi$) 2D MBs as the system length $L$ is varied from being
small to large with respect to $\xi$, hence modulating escape time
relative to $\hbar/\langle \delta\mu \rangle$ and summarize the values
of $L/\xi$ for the systems depicted in
Figs.~\ref{FIG:PWaveStatistics}a-d and \ref{FIG:SWaveStatistics}a-d. In
Fig.~\ref{FIG:UniversalityCrossoverSmallW}, we note the locations of
all of the Figs.~\ref{FIG:PWaveStatistics}a-d and
\ref{FIG:SWaveStatistics}a-d on the $L/\xi$ axis. All of these systems
have one dimension (say, $W$) much smaller than $\xi$. However we
stress that the numerical simulations depicted here do not use this
approximation. The simulations use the full tight-binding version of
the Bogoliubov--de Gennes Hamiltonian (see
Appendix~\ref{SECT:Appendix:TB_Methods}).)
Fig.~\ref{FIG:UniversalityCrossoverSmallW} clearly shows the
universality crossover in these systems.

\begin{figure}[tb]
\centerline{\includegraphics[width=1\linewidth]{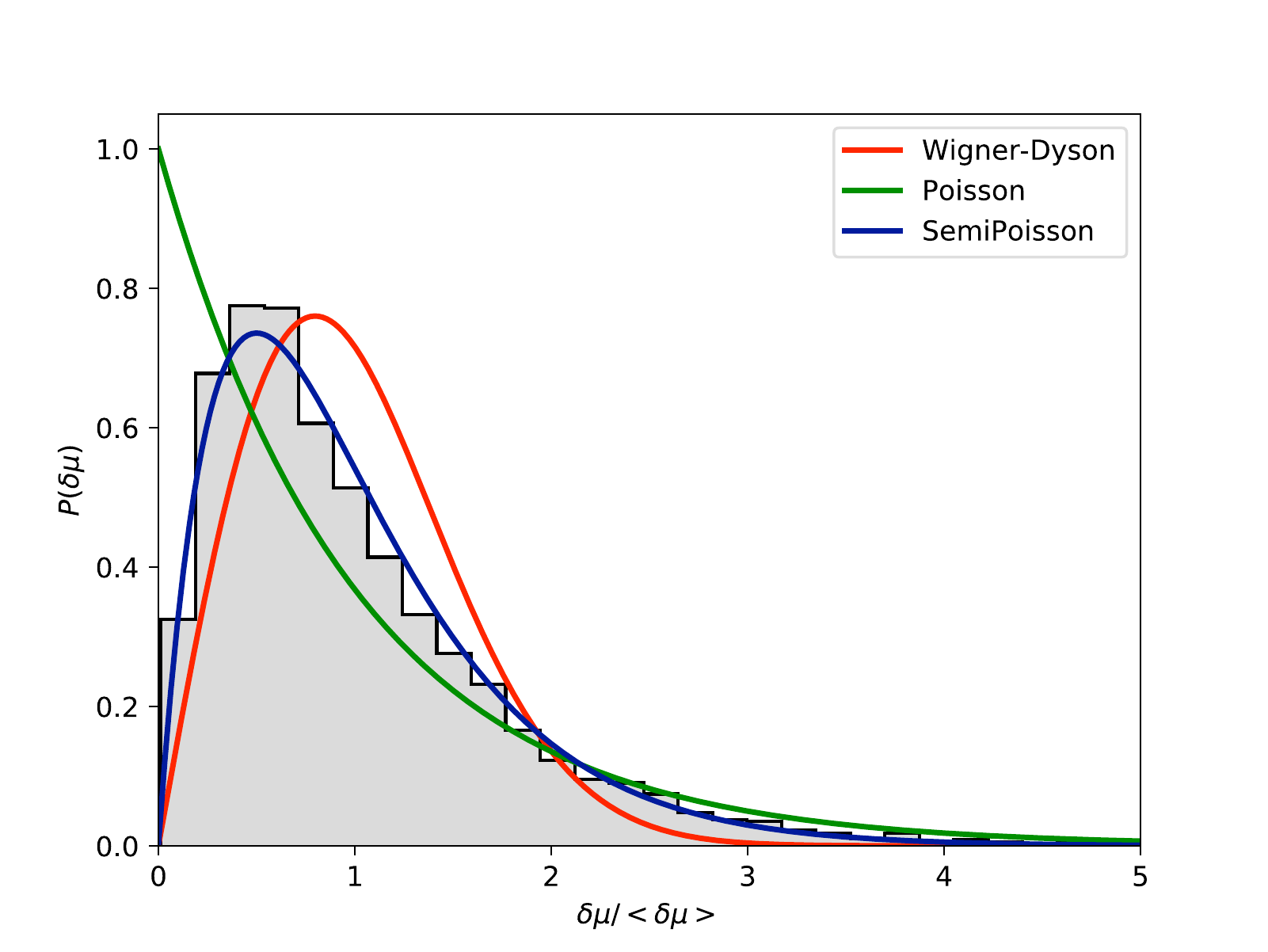}}
\caption{(Color online) Fermion-parity crossing spacing statistics for
a \textit{p}-wave system with \textit{both} dimensions much larger than
$\xi$ ($L=W=5\xi$), showing the statistics obtained from a tight-binding
simulation of a disordered system in a square geometry (500 disorder
realizations) whose parameters are $L=W=80a$, $V_0=0.32t$, $\Delta' =
0.125ta$ and $\xi=16a$.}\label{FIG:LargeSize}
\end{figure}

The short coherence length limit, where the system size exceeds $\xi$
in all directions, was considered by Beenakker \textit{et
al.}~\cite{REF:Beenakker13b} In this case the FPX points have the same
statistics as real eigenvalues of a real {\em non-Hermitian} matrix.
For completeness, we also present the FPX spacing statistics in this
limit in Fig.~\ref{FIG:LargeSize}, where we show the statistics of a
system with both dimensions $L_1$ and $L_2$ much larger than $\xi$,
corresponding to a real Hamiltonian with semi-Poissonian statistics.

%%%%%%%%%%%%%%%%%%%%%%%%%%%%%%%%%%%%%
%									%
%	Figure: L/\xi summary			%
%									%
%%%%%%%%%%%%%%%%%%%%%%%%%%%%%%%%%%%%%

%%%%%%%%%%%%%%%%%%%%%%%%%%%%%%%%%%%%%
%									%
%	Figure: L,W>>\xi				%
%									%
%%%%%%%%%%%%%%%%%%%%%%%%%%%%%%%%%%%%%

%%%%%%%%%%%%%%%%%%%%%%%%%%%%%%%%%%%%%
%									%
%	Conclusions						%
%									%
%%%%%%%%%%%%%%%%%%%%%%%%%%%%%%%%%%%%%

\section{Conclusions}\label{SECT:Conclusions}

In summary we studied the spectra of fermion-parity switches of a
Majorana billiard using methods from semiclassical physics and quantum
chaos. In particular, we show that the average density of
fermion-parity crossings is described by a Weyl expansion and the
disordered billiards feature Lifshitz tails in the fully depleted
limit. Moreover, we demonstrate that the parity crossings has a
tendency to sequentially bunch and anti-bunch, which is reminiscent of
supershell effects in finite systems. We show that the oscillations in
the density of fermion-parity crossings resulting from this bunching
can be obtained by semiclassical means, extending Gutzwiller's trace
formula for conventional quantum billiards to Majorana billiards.
Finally, we show that the fermion-parity crossing spacings obey a
universal distribution as described by random matrix theory. We thus
demonstrate that ``one can hear (information about) the shape of a
Majorana billiard'' from fermion parity switches.

%%%%%%%%%%%%%%%%%%%%%
%					%
% Acknowledgments	%
%					%
%%%%%%%%%%%%%%%%%%%%%

\begin{acknowledgments}

We thank M. Wimmer, K. Richter and C.W.J. Beenakker for useful
discussions. This work was supported by funds of the Erdal
\.{I}n\"{o}n\"{u} chair. {\.I}.A. is a member of the Science
Academy--Bilim Akademisi--Turkey; B.P. and A.M.B thank the Science
Academy--Bilim Akademisi--Turkey for the use of their facilities
throughout this work.

\end{acknowledgments}

%%%%%%%%%%%%%%%%%%%%%%%%%%%%%%%%%%%%%%%%%%%%%%%%
%                                              %
% Appendix                                     %
%                                              %
%%%%%%%%%%%%%%%%%%%%%%%%%%%%%%%%%%%%%%%%%%%%%%%%

\appendix

%%%%%%%%%%%%%%%%%%%%%%%%%%%%%%%%%%%%%%%%%%%%%%%%
%                                              %
% Numerical Simulationss                        %
%                                              %
%%%%%%%%%%%%%%%%%%%%%%%%%%%%%%%%%%%%%%%%%%%%%%%%

\section{Numerical tight-binding simulations}\label{SECT:Appendix:TB_Methods}

In order to demonstrate our analytical results in 
Sections~\ref{SUBSECT:MapToWeylPWave} and \ref{SUBSECT:MapToWeylSWave} 
for average density of fermion-parity crossings, we perform 
tight-binding simulations of fermion-parity crossings in a 
\textit{p}-wave and \textit{s}-wave MBs using the Kwant toolbox for 
quantum transport~\cite{REF:Kwant14}.

For the \textit{p}-wave numerical results, we start with the LHS of
Eq.~(\ref{EQN:ModifiedHamiltonian_PWave}), which is a non-Hermitian
operator, as opposed to the \textit{p}-wave Hamiltonian in
Eq.~(\ref{EQN:Hamiltonian_PWave}). This non-Hermitean operator and the
\textit{p}-wave Hamiltonian in Eq.~(\ref{EQN:Hamiltonian_PWave}) are
equivalent in the sense that no approximation was made in going from
Eq.~(\ref{EQN:Hamiltonian_PWave}) to
Eq.~(\ref{EQN:ModifiedHamiltonian_PWave}). We convert this
non-Hermitean operator to its tight-binding form, which satisfies
$\hat{O}_\mathrm{TB}^\mathrm{PW} \chi = \mu \chi$, using conventional
methods (see, for example, Ref.~[\onlinecite{REF:Book:Datta97}]):
\begin{align}\label{EQN:Appendix:PWaveNonHermitean_TB}
\hat{O}_\mathrm{TB}^\mathrm{PW} &= \big( 2dt + V(x,y)\big)\,\tau_0\,\ket{x,y}\bra{x,y} \nonumber\\
								&\qquad -t \tau_0 \big[\ket{x+a,y}\bra{x,y} + \ket{x,y+a}\bra{x,y} + \mathrm{h.c.}\big] \nonumber\\
								&\qquad +i \Delta' \bigg[ \frac{i}{2} \tau_y \ket{x+a,y}\bra{x,y}\nonumber\\
								&\qquad\qquad - \frac{i}{2} \tau_x \ket{x,y+a}\bra{x,y} + \mathrm{h.c.}\bigg],
\end{align}
where $t = \hbar^2/2 m a^2$ is the hopping parameter, $a$ is the
lattice constant for the tight-binding lattice and $V(x,y)$ is the
onsite potential. For disordered systems, we take the disorder to be
Gaussian, i.e. $\left\langle V(\mathbf{r}) V(\mathbf{r}') \right\rangle
=D \delta(\mathbf{r}-\mathbf{r}')$ for $\mathbf{r}, \mathbf{r}'$ within
the system, where $\left\langle \ldots \right\rangle$ represents
averaging over disorder realizations,  $D \equiv V_d^2 a^d$ with $V_d$
being the disorder strength and $d$ is the dimension of the system.
(In most of our manuscript, $d=2$; if $d=1$, then the hoppings in the
$y$-direction are absent). In tight-binding simulations, this
corresponds to choosing randomly the on-site potential from a Gaussian
distribution. For ballistic cavity results, we set $V(x,y) = 0$ within
the cavity. The boundaries of the system are defined by the lack of
hopping to outside. We form the tight-binding sparse matrix of this
operator using the Kwant library~\cite{REF:Kwant14} over the system
shape described in Fig.~\ref{FIG:2DSystemShape} and the relevant
plots. We then numerically obtain the eigenvalues of this
(non-Hermitian) sparse matrix using LAPACK libraries present in the
SciPy package~\cite{REF:Scipy01}. We finally discard non-real
eigenvalues to obtain our results.

For the \textit{s}-wave results, we go through the same procedure, 
except for utilizing the appropriate tight-binding-representation of 
the non-Hermitian operator derived from the Hamiltonian in 
Eq.~(\ref{EQN:Hamiltonian_SWave}). For $E=0$, the tight-binding model 
for the \textit{s}-wave equivalent of 
Eq.~(\ref{EQN:ModifiedHamiltonian_PWave}) reads 
$\hat{O}_\mathrm{TB}^\mathrm{SW} \chi = \mu \chi$, with the 
non-Hermitian operator $\hat{O}_\mathrm{TB}^\mathrm{SW}$ defined as:
\begin{align}\label{EQN:Appendix:SWaveNonHermitean_TB}
\hat{O}_\mathrm{TB}^\mathrm{SW} &= \big[\big( 2dt + V(x,y)\big)\,\sigma_0 \tau_0 + B \, \sigma_x \tau_z \big]\,
								\ket{x,y}\bra{x,y} \nonumber\\
								&\qquad -t \sigma_0 \tau_0 \big[\ket{x+a,y}\bra{x,y} + \ket{x,y+a}\bra{x,y} + \mathrm{h.c.}\big] \nonumber\\
								&\qquad -\sigma_y \tau_0 \,\big[\frac{i\alpha}{2} \,\ket{x+a,y}\bra{x,y} + \mathrm{h.c.}\big]\nonumber\\
								&\qquad +\sigma_x \tau_0 \,\big[\frac{i\alpha}{2} \,\ket{x,y+a}\bra{x,y} + \mathrm{h.c.}\big]\nonumber\\								
								&\qquad +i \Delta \sigma_0 \tau_y \ket{x,y}\bra{x,y}.
\end{align}
Again, in the plots where $d=1$, the hoppings in the $y$-direction are
absent.

For disorder averaging, we create many realizations of the same
disordered system and do statistics over the combined results of each
realization. For shape averaging over chaotic cavities, we create many
realizations of the same chaotic cavity, the difference between
realizations being the positioning of a relevant geometrical feature of
the cavity, without changing the size of the system volume or boundary.
For the Lorentz cavity, for example, we slightly change the position of
the central stopper for each realization (making sure the stopper never
comes too close to a wall). We check that the change is large enough
numerically to yield a completely different set of eigenvalues.

%%%%%%%%%%%%%%%%%%%%%%%%%%%%%%%%%%%%%%%%%%%%%%%%%%%%%%%
%                                              		  %
% Appendix Section: PWave Shell Effect on a Lattice   %
%                                              		  %
%%%%%%%%%%%%%%%%%%%%%%%%%%%%%%%%%%%%%%%%%%%%%%%%%%%%%%%

\section{Oscillatory behavior of the density of fermion-parity crossings in a
disk Majorana billiard}\label{SECT:Appendix:Oscillations}

\begin{figure}[tb]
\centerline{\includegraphics[width=1\linewidth]{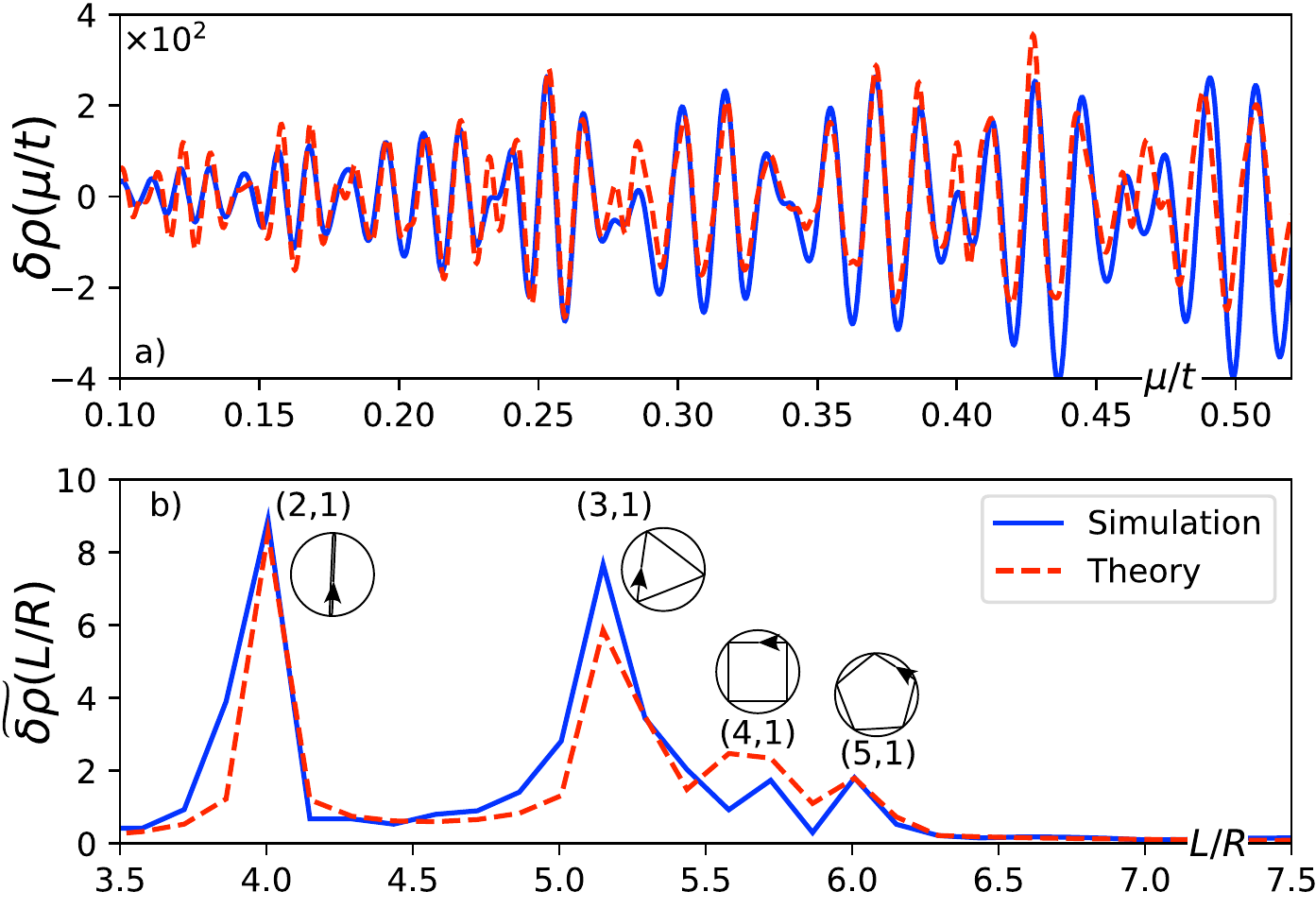}}
\caption{(Color online) a) Density oscillations of fermion-parity
crossings $\rho_\textrm{osc}$ for a clean \textit{p}-wave disk Majorana
billiard on a lattice with $R = 100 a$, $\Delta' = 0.001 ta$. b) The
Fourier transform of $\rho_{osc}$. The $(v, w)$ pairs and corresponding
classical orbits for the peaks are labeled. The smoothing parameter for
both figures is $\gamma =
0.4/R$.}\label{FIG:Appendix:DensityOscillations}
\end{figure}

In this section, we demonstrate the trace formula for $\rho_{\rm osc}$ 
(see Eq.~(\ref{EQN:DOSMeanPlusOsc})) for a \textit{p}-wave disk MB of 
radius $R$. As opposed to the calculation in the main text, here we 
compare the trace formula to tight binding simulations.

We remind the reader that the oscillatory part $\rho_\mathrm{osc}(E)$ 
of the density of states $\rho(E)$ for a two dimensional disk billiard 
of radius $R$ with quadratic dispersion is given 
by~\cite{REF:Book:Brack03}:
\begin{align}\label{EQN:Appendix:GutzwillerDOSCircle}
\rho_\mathrm{osc}(E)	&=	\frac{1}{E_0} \sqrt{\frac{\hbar}{\pi p R}} \,
							\sum_{w=1}^{\infty} \sum_{v=2w}^{\infty}
							f_{vw} \frac{\sin^{3/2}(\varphi_{vw})}{\sqrt{v}}\nonumber\\
					&\quad \times \mathrm{Im}\big[\exp\{i (S_{vw}/\hbar - 3v\pi/2 + 3\pi/4 )\}\big],
\end{align}
with
\begin{align}\label{EQN:Appendix:DiscOrbitLabel}
f_{vw} =	\begin{cases}
			1	& \mbox{if } v = 2w \\
			2  & \mbox{if } v> 2w
			\end{cases}
\end{align}
and $E_0 \equiv \hbar^2/(2mR^2)$. For a quadratic Hamiltonian, $S_{vw} 
= p\, L_{vw} $ is the classical action of the orbit with $L_{vw} = 
2vR\sin(\varphi_{vw})$ being the classical orbit length of 2D disk, 
$\varphi_{vw} \equiv \pi w/v$ is half of the polar angle and $p$ is the 
momentum of the particle. As before, $v, w$ are two integers that 
correspond to the number of vertices and windings of the classical 
periodic orbit, respectively.

However the tight binding dispersion breaks the rotational symmetry of 
the problem weakly. The orbits that belong to the families that have 
the same action for a quadratic dispersion have slightly different 
actions for the tight binding dispersion. This type of symmetry 
breaking can then be treated by the semiclassical perturbation theory 
as discussed in~\cite{REF:Book:Brack03} (see pp. 272). This would 
involve averaging the variation of the phases over all the orientations 
of the orbits, resulting in an effective dispersion $E_\textrm{eff}(p)$ 
of a fictitious rotationally invariant problem. We find that the (one 
dimensional tight-binding--like) dispersion $E_\textrm{eff} = 2t\,(1 - 
\cos{(pa/\hbar)}) $ produces a very good fit to the numerical 
simulations. We thus obtain the expression for momentum $p(\mu)$:
\begin{align}\label{EQN:Appendix:p_TB}
p(\mu) &= \frac{\hbar}{a}\arccos\bigg(1-\frac{\mu}{2t}\bigg).
\end{align}
The deviations from the quadratic dispersion lead to a correction 
$S_{vw} \rightarrow S_{vw} + \Delta S_{vw} $ in the action:
\begin{align}
\label{EQN:Appendix:ActionCorrection}
\Delta S_{vw} = \frac{\hbar}{a}\tan\bigg(\frac{p(\mu) a}{2\hbar}\bigg)L_{vw}.
\end{align}
We now obtain the oscillatory part of the density of fermion-parity 
crossings corrected for tight binding dispersion:
\begin{align}
\rho_\mathrm{osc}(\mu)	&=		\frac{1}{E_0}\, \bigg(\frac{\hbar}{\pi R \, p(\mu)} \bigg)^{1/2}\,
								\sum_{w=1}^{\infty} \sum_{v=2w}^{\infty}
								f_{vw}\frac{\sin^{3/2}(\varphi_{vw})}{\sqrt{v}} \nonumber\\
						&\quad	\times \mathrm{Im}\bigg[\exp\bigg\{i L_{vw}\nonumber\\
						&\qquad \times \, \bigg(\frac{p(\mu+i\gamma)}{\hbar} -\frac{1}{a}\,\tan\frac{p(\mu + i\gamma)\, a}{2\hbar} \bigg) \nonumber\\
						&\quad	+i\big(- 3v\pi/2 + 3\pi/4 \big) \bigg\}\bigg].
\end{align}
Here, we combined Eq.~(\ref{EQN:Appendix:GutzwillerDOSCircle}), 
(\ref{EQN:Appendix:p_TB}) and (\ref{EQN:Appendix:ActionCorrection}) at 
$\mu \rightarrow \mu+i\gamma$, with $\gamma$ being the smoothing 
parameter.

The numerical results for $\rho_\mathrm{osc}$ and 
$\widetilde{\rho}_\mathrm{osc}$ plotted in 
Fig.~\ref{FIG:Appendix:DensityOscillations} is obtained by solving a 
tight-binding \textit{p}-wave system shaped as a disk using the Kwant 
toolbox as described in Appendix~\ref{SECT:Appendix:TB_Methods}. We 
then obtain $\rho_\mathrm{osc}$ as
\begin{align}\label{EQN:Appendix:rho_osc_TB_PW}
\rho_\mathrm{osc}(\mu/t)	&= 	\rho_\gamma(\mu/t) - \rho_\textrm{w}(\mu/t)	,
\end{align}
where $\rho_\textrm{w}$ corresponds to the volume and surface terms of 
the Weyl expansion in Eq.~(\ref{EQN:WeylExpansion_PWave}) and 
$\rho_{\gamma}$ is the smoothed density of fermion-parity crossings
\begin{align}\label{EQN:Appendix:SmoothedSimulatedDensity}
\rho_\gamma(\mu/t)	&=	\int d\mu' \sum_{\mu_c} \delta(\mu'-\mu_c) \,
						F\bigg(\frac{\mu-\mu'}{\gamma}\bigg), \nonumber\\
\end{align}
$F\big(\frac{\mu-\mu'}{\gamma}\big)$ is the Gaussian smoothing function 
with smoothing width $\gamma$. We then take the Fourier transform of 
$\rho_\mathrm{osc}(k(\mu/t)\, a) \xrightarrow{\mathrm{FT}} 
\widetilde{\rho}_\mathrm{osc}(L/R)$ to identify the peaks corresponding 
to the lowest length $L$ and the highest symmetry semiclassical 
periodic orbits~\cite{REF:Book:Brack03} and plot the results in 
Fig.~\ref{FIG:Appendix:DensityOscillations}b. We find good agreement 
with our analytical results.


\begin{thebibliography}{99}
\bibitem{REF:Footnote:MBRealizations}
These topological superconductor systems could be comprised of a 
finite-sized superconductor or a finite-sized normal-state region 
proximity coupled to a superconductor (also known as an Andreev 
billiard~\cite{REF:Kostzin95, REF:Adagideli02a, REF:Beenakker05}).
\bibitem{REF:Footnote:IsospectralDomains}
Although isospectral domains of different shapes
exist~\cite{REF:Gordon92}, it turns out to be possible to extract
geometrical and dynamical information from the energy
spectra~\cite{REF:Book:Weyl68}.
\bibitem{REF:Footnote:3DDoFPX}
We note that the $d=3$ case in Eq.~(\ref{EQN:WeylExpansion_PWave})
is a trivial extension of the $d=2$ case in that the \textit{p}-wave
coupling term is considered to be a 2D coupling.
\end{thebibliography}
\end{document}